\documentclass[a4paper,12pt]{article}
\usepackage[dvips]{graphicx}
\usepackage[dvips,usenames]{color}
\usepackage{amsmath,amssymb}

\def\gev{{\rm GeV}}

\def\etal{{\it et al.}}

\newcommand{\beq}{\begin{equation}}
\newcommand{\eeq}{\end{equation}}
\newcommand{\bea}{\begin{eqnarray}}
\newcommand{\eea}{\end{eqnarray}}
\newcommand{\bsub}{\begin{subequations}}
\newcommand{\esub}{\end{subequations} \noindent}
\newcommand{\clean}{\setcounter{equation}{0}}

\def\PRD#1#2#3{Phys. Rev. D {\bf #1}, #2 (#3)}
\def\NPB#1#2#3{Nucl. Phys. B {\bf #1}, #2 (#3)}

\def\PLB#1#2#3{Phys. Lett. B {\bf #1}, #2 (#3)}
\def\PRL#1#2#3{Phys. Rev. Lett. {\bf #1}, #2 (#3)}

\def\JCAP#1#2#3{JCAP {\bf #1}, #2 (#3)}
\def\JHEP#1#2#3{JHEP {\bf #1}, #2 (#3)}
\begin{document}

\title{The Possible Textures in the Seesaw Realization of the Strong Scaling Ansatz 
and the Implications for Thermal Leptogenesis}
\author{{\bf Midori Obara} \thanks{E-mail: midori@mail.ihep.ac.cn}
\\
{\normalsize \it Institute of High Energy Physics, 
Chinese Academy of Sciences,} \\
{\normalsize \it P.O. Box 918, Beijing 100049, China}
}

\maketitle

\begin{abstract}
\noindent
We classify the textures of the Dirac and the 
right-handed Majorana neutrino mass matrices, $M_D$ and $M_R$, 
which can satisfy the so-called ``Strong Scaling Ansatz'' (SSA) 
within the framework of the seesaw mechanism 
$M_{\nu}=-M_D^T M_R^{-1} M_D$.
We assume that the Dirac neutrino mass matrix has some texture zeros and examine 
which elements should be zero in order to satisfy the SSA, 
by taking into account all possible textures for $M_R$. 
We find that the resulting Dirac neutrino mass matrices have rank 2 
as well as the rank of the effective neutrino mass matrix $M_{\nu}$, or rank 1, 
depending only on the textures of $M_R^{-1}$.
We also consider the three cases of the breaking of the SSA 
by introducing a complex breaking parameter in $M_{\nu}$ 
and show that it can generate the CP violation in the lepton sector as well as 
non-zero $m_3$ and $U_{e3}$.  
We furthermore discuss the implications of the thermal leptogenesis for 
the both cases which satisfy and break the SSA in the basis where $M_R$ is diagonal.
\end{abstract}

\section{Introduction}

Since the discovery of neutrino oscillations by the Super-Kamiokande collaboration, 
solar, atmospheric, reactor and accelerator neutrino experiments 
(Super-Kamiokande~\cite{SK}, SNO~\cite{SNO}, KamLAND~\cite{KM}, K2K~\cite{K2K} 
and MINOS~\cite{MINOS}) have confirmed the evidence of neutrino oscillations.
A global analysis of
current experimental data yields~\cite{valle}
\bea
0.26 \leq & \sin^2 \theta_{12} ~ & \leq 0.40 \; , \nonumber \\
0.34 \leq & \sin^2 \theta_{23} & \leq 0.67 \; , \nonumber \\
& \sin^2 \theta_{13} & \leq 0.050 \; ,
\label{theta12}
\eea
and
\bea
7.1 \times 10^{-5} ~ {\rm eV}^2 \leq & \Delta m_{21}^2 & \leq 8.3
\times
10^{-5} ~ {\rm eV}^2 \; , \nonumber \\
2.0 \times 10^{-3} ~ {\rm eV}^2 \leq & |\Delta m_{32}^2| ~ & \leq
2.8 \times 10^{-5} ~ {\rm eV}^2 \; ,
\label{m31}
\eea
at the $3 \sigma$.

The structures of neutrino mass matrices have been studied in various models 
based on both continuous~\cite{continuous} and 
discrete flavor symmetries~\cite{discrete}, 
and many attempts to connect the flavor 
symmetry approaches to the grand unified thories have been done~\cite{GUT}.
However, these kind of approaches generally receive the corrections from the 
renormalization group effects.
As a new approach independent of the renormalization group effects\footnote{
It has been mentioned that the storong scaling may not be stable 
under radiative corrections in the MSSM 
for large value of $\tan \beta \sim58-60$ in Ref.~\cite{Singh}},
R.N. Mohapatra and W. Rodejohann have recently proposed the strong 
scaling Ansatz (SSA) that the elements of the neutrino mass matrix 
$(M_{\nu})_{\alpha \beta} \equiv m_{\alpha \beta}$ ($\alpha, \beta=e, \mu, \tau$) 
satisfy the following scaling
\bea
\frac{m_{e\mu}}{m_{e\tau}}=\frac{m_{\mu \mu}}{m_{\mu \tau}}
=\frac{m_{\tau \mu}}{m_{\tau \tau}} \equiv c \, ,
\eea
in the basis where the charged lepton mass matrix is diagonal, 
and have shown that such a neutrino mass matrix 
\bea
M_{\nu} =
\left(
\begin{array}{@{\,}ccc@{\,}}
A & B & B/c \\
B & D & D/c \\
B/c & D/c & D/c^2 
\end{array}
\right) 
= U {\rm diag} (m_1,m_2,m_3) U^T \, , 
\eea
where $U$ is the PMNS matrix, predicts the inverted hierarchy with $m_3=0$, 
vanishing $U_{e3}$ and no CP violation~{\cite{MR}}, 
accommodating to the currecnt neutrino experimental data. 
Here we take the following parameterization for $U$~\cite{ParaPMNS}:
\bea
U=
\left(
\begin{array}{@{\,}ccc@{\,}}
c_{12}c_{13} & s_{12}c_{13} & s_{13} e^{-i \delta} \\
-c_{23}s_{12}-s_{23}s_{13}c_{12}e^{i \delta} 
& c_{23}c_{12}-s_{23}s_{13}s_{12}e^{i \delta} 
& s_{23}c_{13} \\
s_{23}s_{12}-c_{23}s_{13}c_{12}e^{i \delta} 
& -s_{23}c_{12}-c_{23}s_{13}s_{12}e^{i \delta} 
& c_{23}c_{13} 
\end{array}
\right) P \, ,
\eea
where $s_{ij} \equiv \sin \theta_{ij}$, $c_{ij} \equiv \cos \theta_{ij}$ and 
$P={\rm diag} (1,e^{i \alpha},e^{i (\beta+\delta)})$ with $\alpha, \beta$ and $\delta$ 
being the Majorana and Dirac phases.
By adjusting the value of the scaling parameter $c$, 
we can obtain the non-maximal atmospheric neutrino mixing angle which may be favored 
in future experiments.
There are three possible cases of breaking the SSA~{\cite{MR}}:
\bea
&&{\rm A1:} \quad
\frac{m_{e\mu}}{m_{e\tau}}=\frac{m_{\mu \mu}}{m_{\mu \tau}}=c \, ,
\quad
\frac{m_{\tau \mu}}{m_{\tau \tau}} =c(1+\epsilon) \, ,
\label{A1breaking}
\\
&&{\rm A2:} \quad
\frac{m_{e\mu}}{m_{e\tau}}=\frac{m_{\tau \mu}}{m_{\tau \tau}}=c \, ,
\quad
\frac{m_{\mu \mu}}{m_{\mu \tau}}=c(1+\epsilon) \, ,
\label{A2breaking}
\\
&&{\rm A3:} \quad
\frac{m_{\mu \mu}}{m_{\mu \tau}}=\frac{m_{\tau \mu}}{m_{\tau \tau}}=c \, ,
\quad
\frac{m_{e\mu}}{m_{e\tau}}=c(1+\epsilon) \, ,
\label{A3breaking}
\eea
and it has been shown that non-zero $m_3$ and $U_{e3}$
can be generated in these three cases~{\cite{MR}}.
In this paper we will show that both real and complex breaking parameter 
can generate the CP violation in the lepton sector as well as 
non-zero $m_3$ and $U_{e3}$.

This paper is organized as follows. In section 2, we classify the textures of the Dirac and the 
right-handed Majorana neutrino mass matrices, $M_D$ and $M_R$, 
which can satisfy the SSA within the framework of the seesaw mechanism 
$M_{\nu}=-M_D^T M_R^{-1} M_D$, and show the conditions of elements in $M_D$ 
for getting the SSA by taking into account all possible textures for $M_R$. 
In this section, we also consider the three cases of breaking the SSA by introducing a complex 
breaking parameter in $M_{\nu}$ and examine which cases of the breaking can be realized 
within the seesaw framework. 
In section 3, we briefly review the phenomenology of the SSA and 
examine the effects of the breaking of the SSA on $m_3$, $U_{e3}$ 
and $J_{{\rm CP}}$ for the three cases A1, A2 and A3, semi-analytically.
Numerical analyses for the original case of the SSA and 
the three cases A1, A2 and A3 will be done in section 4.
In section 5, we discuss the implications of the thermal leptogenesis for 
the both cases which satisfy and break the SSA in the basis where $M_R$ is diagonal. 
Section 6 is devoted to summary.

\section{Classification}
\clean

In this section, 
we classify the textures of $M_D$ and $M_R$,  
which can satisfy the SSA within the framework of the seesaw mechanism 
$M_{\nu}=-M_D^T M_R^{-1} M_D$.
In order to do that, we take the form of $M_D$ as
\bea
M_{D}=
\left(
\begin{array}{@{\,}ccc@{\,}}
a_1 & b_1 & d_1 \\
a_2 & b_2 & d_2 \\
a_3 & b_3 & d_3 
\end{array}
\right) \, ,
\eea 
and find the conditions of elements in $M_D$ for getting the SSA,
by taking into account all possible textures for $M_R$. 
First, we mention the most general condition of the elements in $M_D$.
If $M_D$ is taken to be the following form,
\bea
M_{D}=
\left(
\begin{array}{@{\,}ccc@{\,}}
a_1 & b_1 & b_1/c \\
a_2 & b_2 & b_2/c \\
a_3 & b_3 & b_3/c 
\end{array}
\right) \, ,
\eea
then $M_{\nu}$ satisfies the strong scaling, 
without depending on the textures for $M_R$~{\cite{BMR}}.
Here we assume that $M_D$ has some texture zeros 
and examine which elements should be zero in order to get the SSA.
The Dirac neutrino mass matrices with texture zeros are attractive to relate 
the low energy CP violation in the lepton sector to 
the high energy CP violation necessary for the thermal 
leptogeneis~\cite{tanimoto,Branco,IR,BRS}.
We have listed the conditions of elements in $M_D$ for getting the SSA 
in Tables~\ref{table1},~\ref{table2},~\ref{table3} and~\ref{table4}.
It is obvious that the textures of $M_R$ for the classes $\tilde{F}_1$--$\tilde{F}_7$ 
in Table~\ref{table3} 
can be obtained by exchanging the generation indices of those 
for the classes $F_1$--$F_7$ in Table~\ref{table2}.
Similarly, the classes $G_2$ and $G_3$ can be obtained from 
the classes $G_4$ and $G_5$ in Table~\ref{table4}.
From our classification, 
we have found that the resulting Dirac neutrino mass matrices have rank 2 
as well as the rank of the effective neutrino mass matrix $M_{\nu}$, or rank 1, 
depending only on the textures of $M_R^{-1}$;
the same texture for $M_R^{-1}$ leads to the same conditions for the elements 
in $M_D$.

We also consider the three cases of breaking the SSA 
by introducing a complex breaking parameter $\epsilon$ in $M_{\nu}$.
Here, all parameters are supposed to be complex, and after rephasing, 
we can redefine $c$, $A$, $B$ and $D$ as real and 
take the neutrino mass matrices for the three cases A1, A2 and A3 
given in Eqs. (\ref{A1breaking}), (\ref{A2breaking}) and (\ref{A3breaking}) as
\bea
{\rm A1:} \quad
M_{\nu} &=&
\left(
\begin{array}{@{\,}ccc@{\,}}
A & B e^{i \phi} & B e^{i \phi}/c \\
B e^{i \phi} & D (1+\epsilon) 
& D(1+\epsilon)/c \\
B e^{i \phi}/c & D(1+\epsilon)/c & D/c^2 
\end{array}
\right) \, , 
\label{A1}
\\
{\rm A2:} \quad
M_{\nu} &=&
\left(
\begin{array}{@{\,}ccc@{\,}}
A & B e^{i \phi} & B e^{i \phi}/c \\
B e^{i \phi} & D(1+\epsilon) & D/c \\
B e^{i \phi}/c & D/c & D/c^2 
\end{array}
\right) \, ,
\label{A2}
\\
{\rm A3:} \quad
M_{\nu} &=&
\left(
\begin{array}{@{\,}ccc@{\,}}
A & B e^{i \phi}(1+\epsilon) & B e^{i \phi}/c \\
B e^{i \phi}(1+\epsilon) & D & D/c \\
B e^{i \phi}/c & D/c & D/c^2 
\end{array}
\right) \, ,
\label{A3}
\eea
where $\epsilon \equiv |\epsilon| e^{i \varphi}$ and $|\epsilon| \ll 1$. 
The conditions of elements in $M_D$ for breaking the SSA have also been listed   
in Tables~\ref{table1},~\ref{table2},~\ref{table3} and~\ref{table4}. 
From these tables, we have found that 
in the class $E$ the cases A1 and A2 can be separately realized 
and the case A3 can only appear in combination with the two cases A1 or A2.
In the other classes, 
only the case A3 can be separately realized  
and admixture of all three cases can also be possible.

Next, we will briefly review the phenomenology of the SSA and 
examine the effects of the breaking of the SSA on $m_3$, $U_{e3}$ 
and $J_{{\rm CP}}$ for the three cases A1, A2 and A3, semi-analytically.

\begin{table}
\caption{The texture of $M_R$ and the conditions of elements in $M_D$ 
for getting and breaking the SSA for the class $E$.} 
\vspace{0.3cm}
\setlength{\tabcolsep}{4pt}\footnotesize
\begin{tabular}{|c|c|c|c|} \hline
Class & $M_R$ & Conditions for getting the SSA  & Conditions for breaking the SSA
\\ \hline
$E$ & 
$\left(
\begin{array}{@{\,}ccc@{\,}}
M_1 & 0 & 0 \\
0 & M_2 & 0 \\
0 & 0 & M_3 \end{array}
\right)^{\mathstrut}_{\mathstrut}$ 
& $\begin{array}{@{\,}c@{\,}}
({\rm e1})~a_1=b_1=d_1=0~{\rm and}~b_2=d_2=0 \\
\\
\\
({\rm e2})~a_1=b_1=d_1=0~{\rm and}~b_3=d_3=0 \\
\\
\\
({\rm e3})~a_2=b_2=d_2=0~{\rm and}~b_1=d_1=0 \\
\\
\\
({\rm e4})~a_2=b_2=d_2=0~{\rm and}~b_3=d_3=0 \\
\\
\\
({\rm e5})~a_3=b_3=d_3=0~{\rm and}~b_1=d_1=0 \\ 
\\
\\
({\rm e6})~a_3=b_3=d_3=0~{\rm and}~b_2=d_2=0 
\end{array}$
& $\begin{array}{@{\,}c@{\,}}
b_1 \neq 0~({\rm A2}),d_1 \neq 0~({\rm A1}) \\
b_2 \neq 0~({\rm A2,A3}),d_2 \neq 0~({\rm A1,A3}) \\
\\
b_1 \neq 0~({\rm A2}),d_1 \neq 0~({\rm A1}) \\
b_3 \neq 0~({\rm A2,A3}),d_3 \neq 0~({\rm A1,A3}) \\
\\
b_2 \neq 0~({\rm A2}),d_2 \neq 0~({\rm A1}) \\
b_1 \neq 0~({\rm A2,A3}),d_1 \neq 0~({\rm A1,A3}) \\
\\
b_2 \neq 0~({\rm A2}),d_2 \neq 0~({\rm A1}) \\
b_3 \neq 0~({\rm A2,A3}),d_3 \neq 0~({\rm A1,A3}) \\
\\
b_3 \neq 0~({\rm A2}),d_3 \neq 0~({\rm A1}) \\
b_1 \neq 0~({\rm A2,A3}),d_1 \neq 0~({\rm A1,A3}) \\
\\
b_3 \neq 0~({\rm A2}),d_3 \neq 0~({\rm A1}) \\
b_2 \neq 0~({\rm A2,A3}),d_2 \neq 0~({\rm A1,A3})
\end{array}$ 
\\ \hline
\end{tabular}
\label{table1}
\end{table}

\begin{table}
\caption{The textures of $M_R$ and the conditions of elements in $M_D$ 
for getting and breaking the SSA for the classes $F_1$--$F_7$. 
Here, ``All'' means the admixture of all three cases A1, A2 and A3.} 
\vspace{0.3cm}
\setlength{\tabcolsep}{4pt}\footnotesize
\begin{tabular}{|c|c|c|c|} \hline
Class & $M_R$ & Conditions for getting the SSA  & Conditions for breaking the SSA
\\ \hline
$F_1$ & 
$\left(
\begin{array}{@{\,}ccc@{\,}}
0 & s & 0 \\
s & 0 & 0 \\
0 & 0 & x \end{array}
\right)^{\mathstrut}_{\mathstrut}$ 
& $\begin{array}{@{\,}c@{\,}}
({\rm f1I})~a_1=b_1=d_1=0 \\
\\
({\rm f1II})~a_2=b_2=d_2=0  
\end{array}$
& $\begin{array}{@{\,}c@{\,}}
a_1 \neq 0~({\rm A3}),b_1 \neq 0~({\rm All}),d_1 \neq 0~({\rm All}) \\
\\
a_2 \neq 0~({\rm A3}),b_2 \neq 0~({\rm All}),d_2 \neq 0~({\rm All})
\end{array}$
\\ \hline
$F_2$ & 
$\left(
\begin{array}{@{\,}ccc@{\,}}
0 & s & 0 \\
s & t & 0 \\
0 & 0 & x \end{array}
\right)^{\mathstrut}_{\mathstrut}$ 
& $\begin{array}{@{\,}c@{\,}}
({\rm f2I})~a_1=b_1=d_1=0 \\
\\
({\rm f2II})~a_2=b_2=d_2=0~{\rm and}~b_3=d_3=0  
\end{array}$
& $\begin{array}{@{\,}c@{\,}}
a_1 \neq 0~({\rm A3}),b_1 \neq 0~({\rm All}),d_1 \neq 0~({\rm All}) \\
\\
a_2 \neq 0~({\rm A3}),b_2 \neq 0~({\rm All}),d_2 \neq 0~({\rm All}) \\
b_3 \neq 0~({\rm All}),d_3 \neq 0~({\rm All})
\end{array}$
\\ \hline
$F_3$ & 
$\left(
\begin{array}{@{\,}ccc@{\,}}
0 & s & 0 \\
s & t & u \\
0 & u & x \end{array}
\right)^{\mathstrut}_{\mathstrut}$ 
& $\begin{array}{@{\,}c@{\,}}
({\rm f3I})~a_1=b_1=d_1=0 \\
\\
({\rm f3II})~a_2=b_2=d_2=0~{\rm and}~a_3=b_3=d_3=0  
\end{array}$
& $\begin{array}{@{\,}c@{\,}}
a_1 \neq 0~({\rm A3}),b_1 \neq 0~({\rm All}),d_1 \neq 0~({\rm All}) \\
\\
a_2 \neq 0~({\rm A3}),b_2 \neq 0~({\rm All}),d_2 \neq 0~({\rm All}) \\
a_3 \neq 0~({\rm A3}),b_3 \neq 0~({\rm All}),d_3 \neq 0~({\rm All})
\end{array}$
\\ \hline
$F_4$ & 
$\left(
\begin{array}{@{\,}ccc@{\,}}
0 & s & 0 \\
s & 0 & u \\
0 & u & x \end{array}
\right)^{\mathstrut}_{\mathstrut}$ 
& same as $F_3$
& same as $F_3$
\\ \hline
$F_5$ & 
$\left(
\begin{array}{@{\,}ccc@{\,}}
0 & s & z \\
s & 0 & u \\
z & u & x \end{array}
\right)^{\mathstrut}_{\mathstrut}$ 
& $\begin{array}{@{\,}c@{\,}}
({\rm f5I})~a_1=b_1=d_1=0~{\rm and}~a_2=b_2=d_2=0 \\
\\
\\
({\rm f5II})~a_2=b_2=d_2=0~{\rm and}~a_3=b_3=d_3=0 \\
\\
\\
({\rm f5III})~a_1=b_1=d_1=0~{\rm and}~a_3=b_3=d_3=0
\end{array}$
& $\begin{array}{@{\,}c@{\,}}
a_1 \neq 0~({\rm A3}),b_1 \neq 0~({\rm All}),d_1 \neq 0~({\rm All}) \\
a_2 \neq 0~({\rm A3}),b_2 \neq 0~({\rm All}),d_2 \neq 0~({\rm All}) \\
\\
a_2 \neq 0~({\rm A3}),b_2 \neq 0~({\rm All}),d_2 \neq 0~({\rm All}) \\
a_3 \neq 0~({\rm A3}),b_3 \neq 0~({\rm All}),d_3 \neq 0~({\rm All}) \\
\\
a_1 \neq 0~({\rm A3}),b_1 \neq 0~({\rm All}),d_1 \neq 0~({\rm All}) \\
a_3 \neq 0~({\rm A3}),b_3 \neq 0~({\rm All}),d_3 \neq 0~({\rm All})
\end{array}$
\\ \hline
$F_6$ & 
$\left(
\begin{array}{@{\,}ccc@{\,}}
0 & s & z \\
s & t & u \\
z & u & x \end{array}
\right)^{\mathstrut}_{\mathstrut}$ 
& same as $F_5$
& same as $F_5$
\\ \hline
$F_7$ & 
$\left(
\begin{array}{@{\,}ccc@{\,}}
y & s & z \\
s & t & 0 \\
z & 0 & x \end{array}
\right)^{\mathstrut}_{\mathstrut}$ 
& same as $F_5$
& same as $F_5$
\\ \hline
\end{tabular}
\label{table2}
\end{table}
\begin{table}
\caption{The textures of $M_R$ and the conditions of elements in $M_D$ 
for getting and breaking the SSA for the classes $\tilde{F}_1$--$\tilde{F}_7$.
Here, ``All'' means the admixture of all three cases A1, A2 and A3.} 
\vspace{0.3cm}
\setlength{\tabcolsep}{4pt}\footnotesize
\begin{tabular}{|c|c|c|c|} \hline
Class & $M_R$ & Conditions for getting the SSA  & Conditions for breaking the SSA
\\ \hline
$\tilde{F}_1$ & 
$\left(
\begin{array}{@{\,}ccc@{\,}}
x & 0 & 0 \\
0 & 0 & s \\
0 & s & 0 \end{array}
\right)^{\mathstrut}_{\mathstrut}$ 
& $\begin{array}{@{\,}c@{\,}}
({\rm \tilde{f}1I})~a_2=b_2=d_2=0 \\
\\
({\rm \tilde{f}1II})a_3=b_3=d_3=0  
\end{array}$
& $\begin{array}{@{\,}c@{\,}}
a_2 \neq 0~({\rm A3}),b_2 \neq 0~({\rm All}),d_2 \neq 0~({\rm All}) \\
\\
a_3 \neq 0~({\rm A3}),b_3 \neq 0~({\rm All}),d_3 \neq 0~({\rm All})
\end{array}$
\\ \hline
$\tilde{F}_2$ & 
$\left(
\begin{array}{@{\,}ccc@{\,}}
x & 0 & 0 \\
0 & t & s \\
0 & s & 0 \end{array}
\right)^{\mathstrut}_{\mathstrut}$ 
& $\begin{array}{@{\,}c@{\,}}
({\rm \tilde{f}2I})~a_2=b_2=d_2=0~{\rm and}~b_1=d_1=0 \\
\\
({\rm \tilde{f}2II})~a_3=b_3=d_3=0  
\end{array}$
& $\begin{array}{@{\,}c@{\,}}
a_2 \neq 0~({\rm A3}),b_2 \neq 0~({\rm All}),d_2 \neq 0~({\rm All}) \\
b_1 \neq 0~({\rm All}),d_1 \neq 0~({\rm All}) \\
\\
a_3 \neq 0~({\rm A3}),b_3 \neq 0~({\rm All}),d_3 \neq 0~({\rm All}) \\
\end{array}$
\\ \hline
$\tilde{F}_3$ & 
$\left(
\begin{array}{@{\,}ccc@{\,}}
x & u & 0 \\
u & t & s \\
0 & s & 0 \end{array}
\right)^{\mathstrut}_{\mathstrut}$ 
& $\begin{array}{@{\,}c@{\,}}
({\rm \tilde{f}3I})~a_1=b_1=d_1=0~{\rm and}~a_2=b_2=d_2=0 \\
\\
({\rm \tilde{f}3II})~a_3=b_3=d_3=0  
\end{array}$
& $\begin{array}{@{\,}c@{\,}}
a_1 \neq 0~({\rm A3}),b_1 \neq 0~({\rm All}),d_1 \neq 0~({\rm All}) \\
a_2 \neq 0~({\rm A3}),b_2 \neq 0~({\rm All}),d_2 \neq 0~({\rm All}) \\
\\
a_3 \neq 0~({\rm A3}),b_3 \neq 0~({\rm All}),d_3 \neq 0~({\rm All}) 
\end{array}$
\\ \hline
$\tilde{F}_4$ & 
$\left(
\begin{array}{@{\,}ccc@{\,}}
x & u & 0 \\
u & 0 & s \\
0 & s & 0 \end{array}
\right)^{\mathstrut}_{\mathstrut}$ 
& same as $\tilde{F}_3$
& same as $\tilde{F}_3$
\\ \hline
$\tilde{F}_5$ & 
$\left(
\begin{array}{@{\,}ccc@{\,}}
x & u & z \\
u & 0 & s \\
z & s & 0 \end{array}
\right)^{\mathstrut}_{\mathstrut}$ 
& $\begin{array}{@{\,}c@{\,}}
({\rm \tilde{f}5I})~a_1=b_1=d_1=0~{\rm and}~a_2=b_2=d_2=0 \\
\\
\\
({\rm \tilde{f}5II})~a_2=b_2=d_2=0~{\rm and}~a_3=b_3=d_3=0 \\
\\
\\
({\rm \tilde{f}5III})~a_1=b_1=d_1=0~{\rm and}~a_3=b_3=d_3=0
\end{array}$
& $\begin{array}{@{\,}c@{\,}}
a_1 \neq 0~({\rm A3}),b_1 \neq 0~({\rm All}),d_1 \neq 0~({\rm All}) \\
a_2 \neq 0~({\rm A3}),b_2 \neq 0~({\rm All}),d_2 \neq 0~({\rm All}) \\
\\
a_2 \neq 0~({\rm A3}),b_2 \neq 0~({\rm All}),d_2 \neq 0~({\rm All}) \\
a_3 \neq 0~({\rm A3}),b_3 \neq 0~({\rm All}),d_3 \neq 0~({\rm All}) \\
\\
a_1 \neq 0~({\rm A3}),b_1 \neq 0~({\rm All}),d_1 \neq 0~({\rm All}) \\
a_3 \neq 0~({\rm A3}),b_3 \neq 0~({\rm All}),d_3 \neq 0~({\rm All})
\end{array}$
\\ \hline
$\tilde{F}_6$ & 
$\left(
\begin{array}{@{\,}ccc@{\,}}
x & u & z \\
u & t & s \\
z & s & 0 \end{array}
\right)^{\mathstrut}_{\mathstrut}$ 
& same as $\tilde{F}_5$
& same as $\tilde{F}_5$
\\ \hline
$\tilde{F}_7$ & 
$\left(
\begin{array}{@{\,}ccc@{\,}}
x & 0 & z \\
0 & t & s \\
z & s & y \end{array}
\right)^{\mathstrut}_{\mathstrut}$ 
& same as $\tilde{F}_5$
& same as $\tilde{F}_5$
\\ \hline
\end{tabular}
\label{table3}
\end{table}
\begin{table}
\caption{The textures of $M_R$ and the conditions of elements in $M_D$ 
for getting and breaking the SSA for the classes $G_1$--$G_6$ and $H_1$--$H_3$.
Here, ``All'' means the admixture of all three cases A1, A2 and A3.} 
\vspace{0.3cm}
\setlength{\tabcolsep}{4pt}\footnotesize
\begin{tabular}{|c|c|c|c|} \hline
Class & $M_R$ & Conditions for getting the SSA  & Conditions for breaking the SSA
\\ \hline
$G_1$ &
$\left(
\begin{array}{@{\,}ccc@{\,}}
0 & 0 & z \\
0 & t & 0 \\
z & 0 & 0 \end{array}
\right)^{\mathstrut}_{\mathstrut}$ 
& $\begin{array}{@{\,}c@{\,}}
({\rm g1I})~a_1=b_1=d_1=0 \\
\\
({\rm g1II})~a_3=b_3=d_3=0  
\end{array}$
& $\begin{array}{@{\,}c@{\,}}
a_1 \neq 0~({\rm A3}),b_1 \neq 0~({\rm All}),d_1 \neq 0~({\rm All}) \\
\\
a_3 \neq 0~({\rm A3}),b_3 \neq 0~({\rm All}),d_3 \neq 0~({\rm All})
\end{array}$
\\ \hline
$G_2$ &
$\left(
\begin{array}{@{\,}ccc@{\,}}
0 & s & z \\
s & t & 0 \\
z & 0 & 0 \end{array}
\right)^{\mathstrut}_{\mathstrut}$ 
& $\begin{array}{@{\,}c@{\,}}
({\rm g2I})~a_1=b_1=d_1=0~{\rm and}~a_2=b_2=d_2=0 \\
\\
({\rm g2II})~a_3=b_3=d_3=0  
\end{array}$
& $\begin{array}{@{\,}c@{\,}}
a_1 \neq 0~({\rm A3}),b_1 \neq 0~({\rm All}),d_1 \neq 0~({\rm All}) \\
a_2 \neq 0~({\rm A3}),b_2 \neq 0~({\rm All}),d_2 \neq 0~({\rm All}) \\
\\
a_3 \neq 0~({\rm A3}),b_3 \neq 0~({\rm All}),d_3 \neq 0~({\rm All}) 
\end{array}$
\\ \hline
$G_3$ &
$\left(
\begin{array}{@{\,}ccc@{\,}}
y & s & z \\
s & t & 0 \\
z & 0 & 0 \end{array}
\right)^{\mathstrut}_{\mathstrut}$ 
& same as $G_2$
& same as $G_2$
\\ \hline
$G_4$ &
$\left(
\begin{array}{@{\,}ccc@{\,}}
0 & 0 & z \\
0 & t & s \\
z & s & 0 \end{array}
\right)^{\mathstrut}_{\mathstrut}$ 
& 
$\begin{array}{@{\,}c@{\,}}
({\rm g4I})~a_1=b_1=d_1=0 \\
\\
({\rm g4II})~a_2=b_2=d_2=0~{\rm and}~a_3=b_3=d_3=0 
\end{array}$
& $\begin{array}{@{\,}c@{\,}}
a_1 \neq 0~({\rm A3}),b_1 \neq 0~({\rm All}),d_1 \neq 0~({\rm All}) \\
\\
a_2 \neq 0~({\rm A3}),b_2 \neq 0~({\rm All}),d_2 \neq 0~({\rm All}) \\
a_3 \neq 0~({\rm A3}),b_3 \neq 0~({\rm All}),d_3 \neq 0~({\rm All})
\end{array}$
\\ \hline
$G_5$ &
$\left(
\begin{array}{@{\,}ccc@{\,}}
0 & 0 & z \\
0 & t & s \\
z & s & y \end{array}
\right)^{\mathstrut}_{\mathstrut}$ 
& same as $G_4$
& same as $G_4$
\\ \hline
$G_6$ &
$\left(
\begin{array}{@{\,}ccc@{\,}}
0 & s & z \\
s & t & u \\
z & u & 0 \end{array}
\right)^{\mathstrut}_{\mathstrut}$ 
& $\begin{array}{@{\,}c@{\,}}
({\rm g6I})~a_1=b_1=d_1=0~{\rm and}~a_2=b_2=d_2=0 \\
\\
\\
({\rm g6II})~a_2=b_2=d_2=0~{\rm and}~a_3=b_3=d_3=0 \\
\\
\\
({\rm g6III})~a_1=b_1=d_1=0~{\rm and}~a_3=b_3=d_3=0
\end{array}$
& $\begin{array}{@{\,}c@{\,}}
a_1 \neq 0~({\rm A3}),b_1 \neq 0~({\rm All}),d_1 \neq 0~({\rm All}) \\
a_2 \neq 0~({\rm A3}),b_2 \neq 0~({\rm All}),d_2 \neq 0~({\rm All}) \\
\\
a_2 \neq 0~({\rm A3}),b_2 \neq 0~({\rm All}),d_2 \neq 0~({\rm All}) \\
a_3 \neq 0~({\rm A3}),b_3 \neq 0~({\rm All}),d_3 \neq 0~({\rm All}) \\
\\
a_1 \neq 0~({\rm A3}),b_1 \neq 0~({\rm All}),d_1 \neq 0~({\rm All}) \\
a_3 \neq 0~({\rm A3}),b_3 \neq 0~({\rm All}),d_3 \neq 0~({\rm All})
\end{array}$
\\ \hline
$H_1$  & 
$\left(
\begin{array}{@{\,}ccc@{\,}}
y & s & 0 \\
s & 0 & u \\
0 & u & x \end{array}
\right)^{\mathstrut}_{\mathstrut}$ 
& same as $G_6$
& same as $G_6$
\\ \hline
$H_2$ & 
$\left(
\begin{array}{@{\,}ccc@{\,}}
y & s & 0 \\
s & t & u \\
0 & u & x \end{array}
\right)^{\mathstrut}_{\mathstrut}$ 
& same as $G_6$
& same as $G_6$
\\ \hline
$H_3$ &
$\left(
\begin{array}{@{\,}ccc@{\,}}
y & s & z \\
s & t & u \\
z & u & x \end{array}
\right)^{\mathstrut}_{\mathstrut}$ 
& same as $G_6$
& same as $G_6$
\\ \hline
\end{tabular}
\label{table4}
\end{table}

\section{Neutrino masses and mixing angles in the SSA and 
the effect of the breaking of the SSA}
\clean

Let us decompose the neutrino mass matrices for the cases A1, A2 and A3 as
\bea
M_{\nu}=M_{\nu}^{(0)}+M_{\nu}^{(1)} \, ,
\eea
where
\bea
M_{\nu}^{(0)} &=&
\left(
\begin{array}{@{\,}ccc@{\,}}
A & B e^{i \phi} & B e^{i \phi}/c \\
B e^{i \phi} & D & D/c \\
B e^{i \phi}/c & D/c & D/c^2 \end{array}
\right) \, , 
\label{zeroth}
\eea
and
\bea
{\rm A1:} \quad
M_{\nu}^{(1)} &=&
\left(
\begin{array}{@{\,}ccc@{\,}}
0 & 0 & 0 \\
0 & D |\epsilon| e^{i \varphi} & D |\epsilon| e^{i \varphi}/c \\
0 & D |\epsilon| e^{i \varphi}/c & 0 \end{array}
\right) \, , 
\\
{\rm A2:} \quad
M_{\nu}^{(1)} &=&
\left(
\begin{array}{@{\,}ccc@{\,}}
0 & 0 & 0 \\
0 & D |\epsilon| e^{i \varphi} & 0 \\
0 & 0 & 0 \end{array}
\right) \, , 
\\
{\rm A3:} \quad
M_{\nu}^{(1)} &=&
\left(
\begin{array}{@{\,}ccc@{\,}}
0 & B |\epsilon| e^{i (\phi+\varphi)} & 0 \\
B |\epsilon| e^{i (\phi+\varphi)} & 0 & 0 \\
0 & 0 & 0 \end{array}
\right) \, . 
\eea 
We diagonalize the mass matrix as 
$U_{\nu}^{\dagger} M_{\nu} M_{\nu}^{\dagger} U_{\nu}=
{\rm diag} (m_1^2,m_2^2,m_3^2)$
by using the following decomposition:
\bea
M_{\nu}M_{\nu}^{\dagger}=M_{\nu}^{(0)} M_{\nu}^{(0) \dagger}
+\delta \mathcal{M} \, ,
\label{hermitianmass}
\eea
where
\bea
\delta \mathcal{M} \equiv M_{\nu}^{(0)} M_{\nu}^{(1) \dagger}
+M_{\nu}^{(1) \dagger} M_{\nu}^{(0)} + M_{\nu}^{(1)} M_{\nu}^{(1) \dagger} \, ,
\eea
and then the unitary matrix for diagonalization of Eq. (\ref{hermitianmass}) 
is decomposed as 
\bea
U_{\nu} \equiv U_{\nu}^{(0)}+U_{\nu}^{(1)} \, .
\eea
First, let us diagonalize the unperturbed part of Eq. (\ref{hermitianmass})\footnote{
The diagonalization of Eq.~(\ref{zeroth}) for the case of $c=1$ has been discussed in Ref.~\cite{KT}.}, 
which satisfies the SSA,
\bea
M_{\nu}^{(0)} M_{\nu}^{(0) \dagger} &=&
\left(
\begin{array}{@{\,}ccc@{\,}}
A' & B' e^{i \phi'} & B' e^{i \phi'}/c \\
B' e^{-i \phi'} & D' & D'/c \\
B' e^{-i \phi'}c & D'/c & D'/c^2 \end{array}
\right) 
\nonumber \\
&=& P_{\nu}^{\dagger} 
\left(
\begin{array}{@{\,}ccc@{\,}}
A' & B' & B' /c \\
B' & D' & D'/c \\
B' /c & D'/c & D'/c^2 \end{array}
\right) P_{\nu} \, ,
\label{neutrino-mass0}
\eea
with $P_{\nu} \equiv {\rm diag}(e^{i \phi'},1,1)$ and
\bea
A' &\equiv& A^2+B^2(1+1/c^2) \, , \\
B' e^{i \phi'} &\equiv& ABe^{-i \phi}+BD e^{i \phi}(1+1/c^2) \, , \\
D' &\equiv& B^2+D^2(1+1/c^2) \, .
\eea
The $M_{\nu}^{(0)} M_{\nu}^{(0)\dagger}$ can be diagonalized as 
$U_{\nu}^{(0) \dagger} M_{\nu}^{(0)} M_{\nu}^{(0) \dagger} U_{\nu}^{(0)}=
{\rm diag} ((m_1^2)^{(0)},(m_2^2)^{(0)},(m_3^2)^{(0)})$,
where the mass eigenvalues for Eq. (\ref{neutrino-mass0}) are given by
\bea
(m_1^2)^{(0)} &=& \frac{1}{2} \{ D'(1+1/c^2)+A'-w \} \, , \\
(m_2^2)^{(0)} &=& \frac{1}{2} \{ D'(1+1/c^2)+A'+w \} \, , \\
(m_3^2)^{(0)} &=& 0 \, ,
\eea
with $w \equiv \sqrt{4B'^2(1+1/c^2)+\{ D'(1+1/c^2)-A' \}^2}$ 
and the unitary matrix for diagonalization of Eq. (\ref{neutrino-mass0}) 
is given by
\bea
U_{\nu}^{(0)}=
\left(
\begin{array}{@{\,}ccc@{\,}}
e^{i \phi'} \cos \theta & e^{i \phi'} \sin \theta & 0 \\
-\frac{c \sin \theta}{\sqrt{1+c^2}} & \frac{c \cos \theta}{\sqrt{1+c^2}}  
& -\frac{1}{\sqrt{1+c^2}} \\
-\frac{\sin \theta}{\sqrt{1+c^2}} & \frac{\cos \theta}{\sqrt{1+c^2}} 
& \frac{c}{\sqrt{1+c^2}} 
\end{array}
\right) \, ,
\label{SSAmixingmatrix}
\eea
with 
\bea
\sin \theta &=& \sqrt{\frac{-(m_1^2)^{(0)}+A'}{(m_2^2)^{(0)}-(m_1^2)^{(0)}}} 
=\sqrt{ \frac{-(D'(1+1/c^2)-A'-w)}{2w} } \, ,
\label{sin}
\\
\cos \theta &=& \sqrt{\frac{(m_2^2)^{(0)}-A'}{(m_2^2)^{(0)}-(m_1^2)^{(0)}}}
=\sqrt{ \frac{D'(1+1/c^2)-A'+w}{2w} } \, .
\label{cos}
\eea
As we can see in Eq.~(\ref{SSAmixingmatrix}), taking $c=1$ leads to the exact 
maximal atmospheric neutrino mixing angle. 
By adjusting the value of $c$, 
we can obtain non-maximal one.
In order for $M_{\nu}^{(0)}$ to be consistent with the experimental data for 
$\Delta m_{{\rm sol}}^2/\Delta m_{{\rm atm}}^2$ 
and $\sin \theta_{12}$, 
both conditions (i)$D'(1+1/c^2)+A' \gg \omega$ corresponding to 
$B \gg A,D$ or $A,D \gg B$,
and 
(ii)$4B'^2(1+1/c^2) \gg \{ D'(1+1/c^2)-A' \}^2$ corresponding to 
$A^2 \simeq D^2 (1+1/c^2)^2$ 
should be satisfied. 

Next, we examine the effect of the breaking of the SSA.
As we have already mentioned in introduction,
non-zero $m_3$ and $U_{e3}$ can be generated by the correction of 
breaking parameter $\epsilon$. 
The explicit forms of the mass egenvalues 
and mixing angles up to the next-leading and leading order 
approximation of the diagonalization of Eq. (\ref{neutrino-mass0}) 
can be seen in appendix, respectively.
Up to the order of $\epsilon$,
we can obtain the approximate relations of $m_3$ and $U_{e3}$ between the cases 
A1 and A2 as
\bea
m_3^{{\rm (A1)}} \simeq m_3^{{\rm (A2)}} \, ,
\quad
U_{e3}^{{\rm (A1)}} \simeq U_{e3}^{{\rm (A2)}}/c^2 \, .
\label{m3A1A2}
\eea
From these relations, we can see that for $c \simeq 1$,
the predictions of $m_3$ and $U_{e3}$ for the case A1 is the almost same as those 
for the case A2. 
For $c>1$, as the deviation of $c$ from 1 becomes larger, 
the value of $U_{e3}$ in the case A2 becomes larger than 
that in the case A1 and for $c<1$ vice versa.
On the other hand, the approximate relation of $U_{e3}$ between the cases A3 
and A2 (A1) can be obtained as 
\bea
U_{e3}^{{\rm (A3)}} 
\simeq \frac{A}{D} U_{e3}^{{\rm (A2)}} 
\simeq c^2 \frac{A}{D} U_{e3}^{{\rm (A1)}}
\, ,
\label{Ue3A3A2A1}
\eea
from which we can see that
the value of $U_{e3}$ in the case A3 is larger than those in the cases A1 and A2 
because of the condition (ii).
The Jarlskog parameter can be written as~\cite{Branco}
\bea
J_{CP}=-\frac{{\rm Im}[h_{12} h_{23} h_{31}]}
{\Delta m_{21}^2 \Delta m_{31}^2 \Delta m_{32}^2} \, ,
\eea
with $h=M_{\nu} M_{\nu}^{\dagger}$. 
Up to the order of $\epsilon$, we can obtain
\bea
{\rm A1:} \quad 
{\rm Im}[h_{12} h_{23} h_{31}] &\simeq&
-\frac{B'^2 D^2}{c^4} |\epsilon| \sin (\phi-\varphi)
-\frac{B' D' BD}{c^4} |\epsilon| \sin \varphi  \, ,
\label{IA1}
\\
{\rm A2:} \quad
{\rm Im}[h_{12} h_{23} h_{31}] &\simeq& 
\frac{B'^2 D^2}{c^2} |\epsilon| \sin \phi'
+\frac{B' D' BD}{c^2} |\epsilon| \sin (\phi-2\phi')  \, ,
\label{IA2}
\\
{\rm A3:} \quad
{\rm Im}[h_{12} h_{23} h_{31}] &\simeq&
\frac{B'^2 B^2}{c^2} |\epsilon| \sin \varphi 
-\frac{B' D' AB}{c^2} |\epsilon| \sin (\phi'+\varphi) 
\nonumber \\
&-& \frac{B' D' BD}{c^2} |\epsilon| 
\{ \sin (\phi'-\varphi)-\sin (\phi'-\phi-\varphi) \}  \, ,
\label{IA3}
\eea
which can lead to non-zero $J_{CP}$ in each case, even if $\varphi=0$. 
Note that the form of $J_{CP}$ in the case A2 does not depend on $\varphi$.
On the other hand, we can see that the second term in Eq.~(\ref{IA1}) 
and the first term in Eq.~(\ref{IA3}) vanish in the case of $\varphi=0$. 
We find that the magnitude of $|J_{CP}|$ can be somewhat enhanced 
by the existence of non-zero $\varphi$ in the case A1, 
as we will see in numerical calculations soon later.

In the next section, we will make the numerical analyses for the neutrino masses,
mixing angles, $J_{{\rm CP}}$ and 
the effective mass for the neutrinoless double beta decay $\langle m_{ee} \rangle$ 
in the original case of the SSA and the three cases A1, A2 and A3.

\section{The numerical analysis}
\clean

In this section, we show the numerical results for the original 
case of the SSA and the three cases A1, A2 and A3. 
For the original case of the SSA,
we can restrict the regions of the input parameters $A$, $B$, $D$, $c$, $\phi$
from the experimental data given in Eqs.~(\ref{theta12}) and (\ref{m31}) 
and determine the value of the effective mass for the neutrinoless double beta decay 
$\langle m_{ee} \rangle$.
In addtion to these five parameters, for the cases A1, A2 and A3,  
we have two more ones $|\epsilon|$ and $\varphi$, which allow us to determine 
the values of $U_{e3}$, $m_3$ and $J_{CP}$. 

In Table~\ref{table5},
we have listed the predicted values of $|U_{e3}|$, $m_3$, $J_{CP}$ and 
$\langle m_{ee} \rangle$ together with the allowed value of $c$, 
for the original case of the SSA and the cases A1, A2 and A3. 
For the original case of the SSA, 
we have considered the two cases of $\phi=0$ and $\phi = 0 \sim 2 \pi$. 
For the cases A1, A2 and A3, the two cases of $\phi = 0 \sim 2 \pi,\varphi=0$ 
and $\phi,\varphi = 0 \sim 2 \pi$ have been considered.  
In Table~\ref{table5}, 
we can see that the maximum value of $|U_{e3}|$ in the case A2 
is larger than that in the case A1.
This is responsible for the deviation of $c$ from 1 in the region of $c>1$.
As seen in Eqs.~(\ref{m3A1A2}) and (\ref{Ue3A3A2A1}), 
the prediction of $m_3$ for the case A1 is the almost same as those for the case A2
and 
the value of $|U_{e3}|$ in the case A3 is larger than those in the cases A1 and A2.
As we have described in the previous section,
we can also see that the magnitude of $|J_{CP}|$ can be somewhat enhanced 
by the existence of non-zero $\varphi$ in the case A1. 
It is in principle possible to detect $|J_{CP}| \sim \mathcal{O}(10^{-2})$  
in the future long-baseline neutrino oscillation experiments.
On the other hand, such a enhancement cannot be seen in the cases A2 and A3. 
Also, the effect of the breaking of the SSA on $\langle m_{ee} \rangle$ cannot be seen.

Because of the texture zeros for $M_D$ in our model,
within the framework of the seesaw mechanism,
we can expect that the predictions of the low energy observables
can be constrained from the baryon asymmetry of the universe through the 
thermal leptogenesis scenario~\cite{FuYa}.
In the next section, we will study the the baryon asymmetry of 
the universe based on the thermal leptogenesis scenario.

\begin{table}
\caption{The predicted values of $|U_{e3}|$, $m_3$, $J_{CP}$ and 
$\langle m_{ee} \rangle$ together with the allowed value of $c$ 
for the original case of the SSA 
and the cases A1, A2 and A3. Here we take $|\epsilon|=0\sim0.25$.} 
\vspace{0.3cm}
\setlength{\tabcolsep}{4pt}\footnotesize
\begin{tabular}{|c|c|c|c|c|} \hline
& SSA $(\phi=0)$ & A1 $(\phi = 0 \sim 2 \pi, \varphi=0)$ 
& A2 $(\phi = 0 \sim 2 \pi, \varphi=0)$ & A3 $(\phi = 0 \sim 2 \pi, \varphi=0)$  
\\ \hline
$c$ & $0.72\sim1.4$ & $0.66\sim1.4$ & $0.68\sim1.4$ & $0.67\sim1.4$
\\ \hline
$|U_{e3}|$ & 0 & $\leq 0.022$ & $\leq 0.026$ & $\leq 0.050$
\\ \hline
$m_3~({\rm eV})$ & 0 & $\leq 2.5 \times 10^{-3}$ 
& $\leq 2.6 \times 10^{-3}$ & $\leq 2.0 \times 10^{-4}$
\\ \hline
$J_{CP}$ & 0 & $-0.0046\sim0.0048$ & $-0.0061\sim0.0054$ & $-0.011\sim0.011$
\\ \hline
$\langle m_{ee} \rangle~({\rm eV})$ & $0.0086\sim0.047$ & $0.095\sim0.052$ 
& $0.010\sim0.052$ & $0.010\sim0.052$
\\ \hline \hline
& SSA $(\phi = 0 \sim 2 \pi)$ & A1 $(\phi, \varphi = 0 \sim 2 \pi)$ 
& A2 $(\phi, \varphi = 0 \sim 2 \pi)$ & A3 $(\phi, \varphi = 0 \sim 2 \pi)$  
\\ \hline
$c$ & $0.74\sim1.4$ & $0.71\sim1.5$ & $0.73\sim1.5$ & $0.72\sim1.5$
\\ \hline
$|U_{e3}|$ & 0 & $\leq 0.031$ & $\leq 0.039$ & $\leq 0.058$
\\ \hline
$m_3~({\rm eV})$ & 0 & $\leq 3.1 \times 10^{-3}$ 
& $\leq 3.0 \times 10^{-3}$ & $\leq 3.3 \times 10^{-4}$
\\ \hline
$J_{CP}$ & 0 & $-0.0047\sim0.0070$ & $-0.0049\sim0.0067$ & $-0.011\sim0.011$
\\ \hline
$\langle m_{ee} \rangle~({\rm eV})$ & $0.011\sim0.051$ & $0.0095\sim0.052$ 
& $0.0099\sim0.053$ & $0.0099\sim0.053$
\\ \hline
\end{tabular}
\label{table5}
\end{table}

\section{Thermal leptogenesis}
\clean
 
In the thermal leptogenesis scenario~\cite{FuYa}, a lepton asymmetry is generated by 
the CP violating out-of-equilibrium decay of heavy right-handed 
Majorana neutrinos $N_i$. Recently, it has been pointed out that the charged lepton 
flavor effects play a crucial role on the dynamics of the thermal leptogenesis 
below the temperature $T \sim M_1 \sim 10^{12}~\gev$~\cite{flavorlepto}. 
For $10^9~\gev \lesssim T \sim M_1 \lesssim 10^{12}~\gev$ and 
for $T \sim M_1 \lesssim 10^9~\gev$,
the interactions mediated by the $\tau$ and $\mu$ are non-negligible.
Thus, the baryon asymmetry should be calculated 
by taking into account the flavor effects.
Considering the flavor effects, 
the CP asymmetry parameter $\epsilon_i^{\alpha}$ is defined as~\cite{flavorlepto}
\bea
\epsilon_i^{\alpha} &\equiv& 
\frac{\Gamma(N_i \to H L_{\alpha})-\Gamma(N_i \to \bar{H} \bar{L}_{\alpha})}
{\Gamma(N_i \to H L_{\alpha})+\Gamma(N_i \to \bar{H} \bar{L}_{\alpha})} 
\nonumber \\
&=& \frac{1}{8\pi v^2} \frac{1}{(M_D M_D^{\dagger})_{ii}} 
\sum_{j \neq i} 
{\rm Im} [(M_D)_{i \alpha} (M_D^{\dagger})_{\alpha j} (M_D M_D^{\dagger})_{ij}] 
f(M_j^2/M_i^2) \, ,
\eea
where $v$ is a vacuum expectation value of the electroweak symmetry breaking 
$v\simeq174~\gev$ and 
\bea
f(x) \equiv \sqrt{x} \Biggl\{ 1-(1+x) \ln \frac{1+x}{x} 
+\frac{1}{1-x}
\Biggr\} \, ,
\label{function}
\eea
with $x \equiv M_j^2/M_i^2$.  
At the temperature $T \sim M_1>10^{12}~\gev$, 
all the charged leptons are out of equilibrium 
and the flavor effects are indistinguishable. 
In this paper, we assume $M_3>M_2>M_1>10^{12}~\gev$  
and thus one flavor approximation is valid~\footnote{
In the one flavor approximation,
for the hierarchical right-handed Majorana neutrinos, 
the lower bound on $M_1$, $M_1>4.9\times 10^{8}\ {\rm GeV}$, 
has been known~\cite{M1bound}.}.
In this temperature regime, 
the CP asymmetry parameter $\epsilon_i$ is given by~\cite{FuYa,lepto}
\bea
\epsilon_i &\equiv& 
\frac{\Gamma(N_i \to H L)-\Gamma(N_i \to \bar{H} \bar{L})}
{\Gamma(N_i \to H L)+\Gamma(N_i \to \bar{H} \bar{L})} 
\nonumber \\
&=& \frac{1}{8\pi v^2} \frac{1}{(M_D M_D^{\dagger})_{ii}} 
\sum_{j \neq i} 
{\rm Im} [((M_D M_D^{\dagger})_{ij}^2] 
f(M_j^2/M_i^2) \, ,
\eea
where $f(x)$ is given in Eq.~(\ref{function}).
In order to calculate the baryon asymmetry of the universe, 
we need to solve the Boltzmann equations~\cite{pu}.
Here we use the approximate solution of the Boltzmann equations as~\cite{BBP}
\bea
\eta_B \simeq 
0.0096 \, \sum_i \epsilon_i \kappa_i \, ,  
\eea
where $\eta_B$ is the baryon asymmetry of the universe and 
$\kappa_i$ is the so-called dilution factor, which describes the wash-out effect 
of the generated lepton asymmetry and is approximated as~\cite{Nielsen}
\bea
\kappa_i \simeq 
0.3 \left(  
\frac{ 10^{-3}~{\rm eV} }{ \tilde{m}_i }
\right)
\left(  
\ln \frac{ \tilde{m}_i }{ 10^{-3}~{\rm eV} } \right)^{-0.6} \, ,
\quad
\tilde{m}_i \equiv \frac{ (M_D M_D^{\dag})_{ii} }{M_i} \, . 
\eea

In this study, we concentrate on the class $E$ with the six realization 
conditions (e1)--(e6) in Table~\ref{table1}, where $M_R$ is diagonal.
Under the six conditions (e1)--(e6), 
we have listed the forms of $M_D, M_D M_D^{\dagger}$ and the CP asymmetry parameter 
$\epsilon_i$ in Table~\ref{table6}. 
Here we denote the class $E$ with the condition (e1) as the case $E$e1 and so on. 
As we can see in Table~\ref{table6}, the forms of $M_D, M_D M_D^{\dagger}$ and $\epsilon_i$ 
in the cases $E$e3 and $E$e5 ($E$e4 and $E$e6) can be obtained 
from those in the case $E$e1 ($E$e2) by relabeling 
the indices of generations as $2 \to 1$ and $2 \to 1, 3 \to 2$, respectively. 
Thus, the pysical consequances for the CP asymmetry in the cases 
$E$e$i$ with $i=1,3,5$ and with $i=2,4,6$ are preserved, respectively. 
As typical examples, we consider the cases $E$e5 and $E$e6.
In Table~\ref{table7}, 
we have listed the predicted values of $\sin^2 \theta_{23}$, $\sin^2 \theta_{12}$, 
$\Delta m^2_{32}$, $\Delta m^2_{21}$ and $\langle m_{ee} \rangle$ 
for the cases $E$e5 and $E$e6.
Here we take $M_2=5 \times 10^{14} \sim 5 \times 10^{15}~\gev$ and
$M_1=(0.01 \sim 0.1) \times M_2$ for the case $E$e5.
In order to be consistent with the experimental data, 
which are given in Eqs.~(\ref{theta12}) and~(\ref{m31}),
and the observed value of $\eta_B$, $\eta_B=(5.9-6.3) \times 10^{-10}$~\cite{WMAP},
we need to take the parameters in $M_D$ as
$a_1\sim \mathcal{O}(1)$, 
$a_2 \sim \mathcal{O}(0.01)$ and $b_2,d_2 \sim \mathcal{O}(1)$.
On the other hand, for the case $E$e6, 
we take $M_2=5 \times 10^{15} \sim 5 \times 10^{16}~\gev$ and
$M_1=(0.01 \sim 0.1) \times M_2$.
Then, the experimental data force the parameters in $M_D$ to be 
$a_1 \sim \mathcal{O}(0.01)$, $a_2 \sim \mathcal{O}(10)$ and 
$b_1,d_1 \sim \mathcal{O}(1)$. 
Thanks of the texture zeros for $M_D$, we can obtain the correlation between
the low energy observable in the lepton sector and 
the high energy CP violation necessary for the thermal leptogeneis; 
the value of $\langle m_{ee} \rangle$ can be 
constrained from $\eta_B$ as  
$0.045~{\rm eV} \lesssim \langle m_{ee} \rangle \lesssim 0.052~{\rm eV}$ 
in the both cases $E$e5 and $E$e6. 

\begin{figure}
\begin{center}
\includegraphics[width=10.0cm,clip]{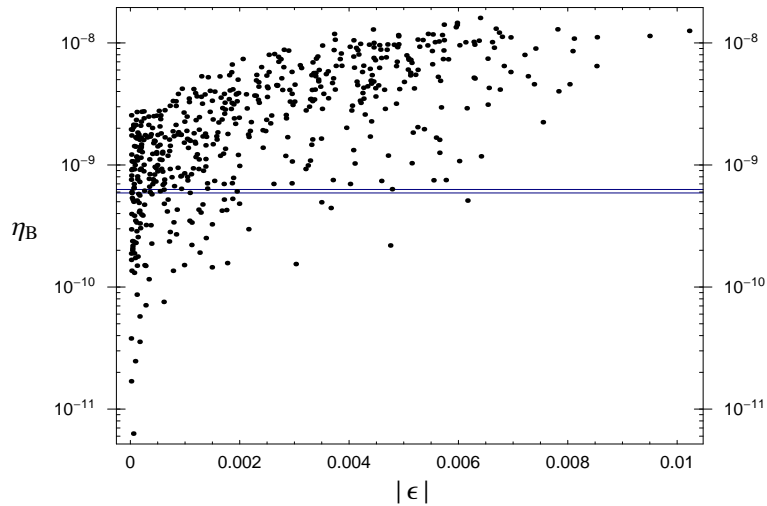}
\end{center}
\caption{The predicted baryon asymmetry as a function of $|\epsilon|$ 
in the case $E$e6 with $b_3 \neq 0$.}
\label{Figure1}
\end{figure}

Finally, we discuss the baryon asymmetry of 
the universe in the case of the breaking of the SSA.
For the case $E$e5 with $b_3 \neq 0$ 
(corresponding to the case A2), 
we can obtain the forms of $M_D$ and $M_D M_D^{\dagger}$ as follows:
\bea
M_D = 
\left(
\begin{array}{@{\,}ccc@{\,}}
a_1 e^{i \xi} & 0 & 0 \\
a_2 & b_2 & d_2 \\
0 & b_3 e^{i \xi'} & 0 \end{array}
\right) v \, , 
M_D M_D^{\dagger} =
\left(
\begin{array}{@{\,}ccc@{\,}}
a_1^2 & a_1 a_2 e^{i \xi} & 0 \\
a_1 a_2 e^{-i \xi} & a_2^2+b_2^2+d_2^2 & b_2 b_3 e^{-i \xi'} \\
0 & b_2 b_3 e^{i \xi'} & b_3^2 \end{array}
\right) v^2 \, .
\eea
From the above forms, we have found that the correction from the breaking of the SSA 
does not affect on $\epsilon_1$\footnote{
This statement holds for the case $E$e5 
with $d_3 \neq 0$ (corresponding to the case A1), because 
we can obtain the form of $M_D M_D^{\dagger}$ 
by rewriting $b_2$ and $b_3$ as $d_2$ and $d_3$ in the 2-3 (3-2) and 3-3 elements, 
respectively.}.
For the case $E$e6 with $b_3 \neq 0$ 
(corresponding to the case A2), we have
\bea
M_D =
\left(
\begin{array}{@{\,}ccc@{\,}}
a_1 & b_1 & d_1 \\
a_2 e^{i \xi} & 0 & 0 \\
0 & b_3 e^{i \xi'} & 0 \end{array}
\right) v \, ,
M_D M_D^{\dagger} =
\left(
\begin{array}{@{\,}ccc@{\,}}
a_1^2+b_1^2+d_1^2 & a_1 a_2 e^{-i \xi} & b_1 b_3 e^{-i \xi'} \\
a_1 a_2 e^{i \xi} & a_2^2 & 0 \\
b_1 b_3 e^{-i \xi'} & 0 & b_3^2 \end{array}
\right) v^2 \, ,
\eea
which leads to the correction for $\epsilon_1$ as
\bea
\Delta \epsilon_1 = \frac{-1}{8\pi} \frac{b_1^2 b_3^2}{a_1^2+b_1^2+d_1^2}
\sin (2\xi') f(M_3^2/M_1^2) \, .
\eea
From the constraint of $\eta_B$, we have found that 
the value of $b_3$ should be of the order of 
$\mathcal{O}(0.1)-\mathcal{O}(1)$.
In Figure.~\ref{Figure1},
we show the predicted value of $\eta_B$ as a function of $|\epsilon|$.
Here we take the values of three right-handed Majorana masses $M_i$ $(i=1,2,3)$ 
as $ M_2=5 \times 10^{15} \sim 5 \times 10^{16}~\gev,
M_1=(0.01 \sim 0.1) \times M_2$ and  
$M_3=(2 \sim 10) \times M_2$.
As we can see in Figure.~\ref{Figure1}, the order of magnitude of the breaking parameter 
$|\epsilon|$ in this case is 
given as $|\epsilon|=(b_3^2/b_1^2) \times (M_1/M_3) \simeq \mathcal{O}(10^{-5})
\sim \mathcal{O}(10^{-3})$\footnote{
For the case $E$e6 with $d_3 \neq 0$ (corresponding to the case A1),  
we can obtain the form of $M_D M_D^{\dagger}$ 
by rewriting $b_1$ and $b_3$ as $d_1$ and $d_3$ in the 1-3 (3-1) and 3-3 elements, 
respectively. Then, we can obtain $\Delta \epsilon_1=(-1/8\pi) 
(d_1^2 d_3^2) \sin (2\xi') f(M_3^2/M_1^2)/(a_1^2+b_1^2+d_1^2)$. 
Because of $b_1,d_1 \sim \mathcal{O}(1)$, $d_3$ should also be of the order of 
$\mathcal{O}(0.1)-\mathcal{O}(1)$ which leads to 
$|\epsilon| =(d_3^2/d_1^2) \times (M_1/M_3)
\simeq \mathcal{O}(10^{-5}) \sim \mathcal{O}(10^{-3})$.},
which can only generate the very small values of 
$m_3 \sim \mathcal{O}(10^{-9})-\mathcal{O}(10^{-6})$, 
$|U_{e3}| \sim \mathcal{O}(10^{-8})-\mathcal{O}(10^{-6})$ and 
$J_{CP} \sim \mathcal{O}(10^{-9})-\mathcal{O}(10^{-7})$. 
Thus, observation of $|U_{e3}|$ in the next generation of reactor and long-baseline 
neutrino experiments 
will exclude the case $E$e6 with $b_3 \neq 0$.

For the case $E$e5 with $b_1 \neq 0$ and 
the case $E$e6 with $b_2 \neq 0$
(corresponding to the combination of the cases A2 and A3), we have
\bea
M_D &=& 
\left(
\begin{array}{@{\,}ccc@{\,}}
a_1 e^{i \xi} & b_1 e^{i \xi'} & 0 \\
a_2 & b_2 & d_2 \\
0 & 0 & 0 \end{array}
\right) v \, ,  
\nonumber \\ 
M_D M_D^{\dagger} &=&
\left(
\begin{array}{@{\,}ccc@{\,}}
a_1^2+b_1^2 & a_1 a_2 e^{i \xi}+b_1 b_2 e^{i \xi'} & 0 \\
a_1 a_2 e^{-i \xi}+b_1 b_2 e^{-i \xi'} & a_2^2+b_2^2+d_2^2 & 0 \\
0 & 0 & 0 \end{array}
\right) v^2 \, ,
\eea
and
\bea
M_D &=& 
\left(
\begin{array}{@{\,}ccc@{\,}}
a_1 & b_1 & d_1 \\
a_2 e^{i \xi} & b_2 e^{i \xi'} & 0 \\
0 & 0 & 0 \end{array}
\right) v \, , 
\nonumber \\
M_D M_D^{\dagger} &=&
\left(
\begin{array}{@{\,}ccc@{\,}}
a_1^2+b_1^2+d_1^2 & a_1 a_2 e^{-i \xi}+b_1 b_2 e^{-i \xi'} & 0 \\
a_1 a_2 e^{i \xi}+b_1 b_2 e^{i \xi'}  & a_2^2+b_2^2 & 0 \\
0 & 0 & 0 \end{array}
\right) v^2 \, ,
\eea
from which we can see that the correction from the breaking of the SSA 
does not affect on $\epsilon_1$ in the both cases
\footnote{Similarly to the case $E$e5 with $b_3 \neq 0$, 
this statement holds for the case $E$e5 ($E$e6)
with $b_1 \neq 0$ ($b_2 \neq 0$) corresponding to the combination of the cases A1 
and A3, 
because we can obtain the form of $M_D M_D^{\dagger}$ 
by rewriting $b_1$ and $b_2$ as $d_1$ and $d_2$ in the 1-2 (2-1) and 2-2 elements, 
respectively.} as well as the $E$e5 with $b_3 \neq 0$.

\begin{table}
\caption{The forms of $M_D$, $M_D M_D^{\dagger}$ and the 
CP asymmetry parameter $\epsilon_i$ for the class $E$ 
with the six conditions (e1)--(e6). 
Here, the case $E$e1 means the class $E$ with the condition (e1) and so on.} 
\vspace{0.3cm}
\setlength{\tabcolsep}{4pt}\footnotesize
\begin{tabular}{|c|c|c|c|} \hline
Case & $M_D/v$ & $M_D M_D^{\dagger}/v^2$ & $\epsilon_i$  
\\ \hline
$E$e1 & 
$\left(
\begin{array}{@{\,}ccc@{\,}}
0 & 0 & 0 \\
a_2 e^{i \xi} & 0 & 0 \\
a_3 & b_3 & d_3 \end{array}
\right)^{\mathstrut}_{\mathstrut}$
&
$\left(
\begin{array}{@{\,}ccc@{\,}}
0 & 0 & 0 \\
0 & a_2^2 & a_2 a_3 e^{i \xi} \\
0 & a_2 a_3 e^{-i \xi} & a_3^2+b_3^2+d_3^2 \end{array}
\right)^{\mathstrut}_{\mathstrut}$ 
& $\begin{array}{@{\,}c@{\,}}
\epsilon_2=\frac{1}{8\pi_{\mathstrut}} a_3^2 \sin (2\xi) f(M_3^2/M_2^2) \\
\epsilon_3=\frac{-1}{8\pi} \frac{a_2^2 a_3^2}{a_3^2+b_3^2+d_3^2} 
\sin (2\xi) f(M_2^2/M_3^2) 
\end{array}$
\\ \hline
$E$e2 &
$\left(
\begin{array}{@{\,}ccc@{\,}}
0 & 0 & 0 \\
a_2 & b_2 & d_2 \\
a_3 e^{i \xi} & 0 & 0 \end{array}
\right)^{\mathstrut}_{\mathstrut}$
&
$\left(
\begin{array}{@{\,}ccc@{\,}}
0 & 0 & 0 \cr
0 & a_2^2+b_2^2+d_2^2 & a_2 a_3 e^{-i \xi} \cr
0 & a_2 a_3 e^{i \xi} & a_3^2 \end{array}
\right)^{\mathstrut}_{\mathstrut}$ 
& $\begin{array}{@{\,}c@{\,}}
\epsilon_2=\frac{-1}{8\pi} \frac{a_2^2 a_3^2}{a_2^2+b_2^2+d_2^2} 
\sin (2\xi) f(M_3^2/M_2^2)
\\
\epsilon_3=\frac{1^{\mathstrut}}{8\pi} a_2^2 \sin (2\xi) f(M_2^2/M_3^2) 
\end{array}$
\\ \hline
$E$e3 &
$\left(
\begin{array}{@{\,}ccc@{\,}}
a_1 e^{i \xi} & 0 & 0 \\
0 & 0 & 0 \\
a_3 & b_3 & d_3 \end{array}
\right)^{\mathstrut}_{\mathstrut}$
&
$\left(
\begin{array}{@{\,}ccc@{\,}}
a_1^2 & 0 & a_1 a_3 e^{i \xi} \\
0 & 0 & 0 \\
a_1 a_3 e^{-i \xi} & 0 & a_3^2+b_3^2+d_3^2 \end{array}
\right)^{\mathstrut}_{\mathstrut}$ 
& $\begin{array}{@{\,}c@{\,}}
\epsilon_1=\frac{1}{8\pi_{\mathstrut}} a_3^2 \sin (2\xi) f(M_3^2/M_1^2) 
\\
\epsilon_3=\frac{-1}{8\pi} \frac{a_1^2 a_3^2}{a_3^2+b_3^2+d_3^2} 
\sin (2\xi) f(M_1^2/M_3^2) 
\end{array}$
\\ \hline
$E$e4 &
$\left(
\begin{array}{@{\,}ccc@{\,}}
a_1 & b_1 & d_1 \\
0 & 0 & 0 \\
a_3 e^{i \xi} & 0 & 0 \end{array}
\right)^{\mathstrut}_{\mathstrut}$
&
$\left(
\begin{array}{@{\,}ccc@{\,}}
a_1^2+b_1^2+d_1^2 & 0 & a_1 a_3 e^{-i \xi} \\
0 & 0 & 0 \\
a_1 a_3 e^{i \xi} & 0 & a_3^2 \end{array}
\right)^{\mathstrut}_{\mathstrut}$ 
& $\begin{array}{@{\,}c@{\,}}
\epsilon_1=\frac{-1}{8\pi} \frac{a_1^2 a_3^2}{a_1^2+b_1^2+d_1^2} 
\sin (2\xi) f(M_3^2/M_1^2)
\\
\epsilon_3=\frac{1^{\mathstrut}}{8\pi} a_1^2 \sin (2\xi) f(M_1^2/M_3^2) 
\end{array}$
\\ \hline
$E$e5 &
$\left(
\begin{array}{@{\,}ccc@{\,}}
a_1 e^{i \xi} & 0 & 0 \\
a_2 & b_2 & d_2 \\
0 & 0 & 0 \end{array}
\right)^{\mathstrut}_{\mathstrut}$
&
$\left(
\begin{array}{@{\,}ccc@{\,}}
a_1^2 & a_1 a_2 e^{i \xi} & 0 \\
a_1 a_2 e^{-i \xi} & a_2^2+b_2^2+d_2^2 & 0 \\
0 & 0 & 0 \end{array}
\right)^{\mathstrut}_{\mathstrut}$ 
& $\begin{array}{@{\,}c@{\,}}
\epsilon_1=\frac{1}{8\pi_{\mathstrut}} a_2^2 \sin (2\xi) f(M_2^2/M_1^2) 
\\
\epsilon_2=\frac{-1}{8\pi} \frac{a_1^2 a_2^2}{a_2^2+b_2^2+d_2^2} 
\sin (2\xi) f(M_1^2/M_2^2) 
\end{array}$
\\ \hline
$E$e6 &
$\left(
\begin{array}{@{\,}ccc@{\,}}
a_1 & b_1 & d_1 \\
a_2 e^{i \xi} & 0 & 0 \\
0 & 0 & 0 \end{array}
\right)^{\mathstrut}_{\mathstrut}$
&
$\left(
\begin{array}{@{\,}ccc@{\,}}
a_1^2+b_1^2+d_1^2 & a_1 a_2 e^{-i \xi} & 0 \\
a_1 a_2 e^{i \xi} & a_2^2 & 0 \\
0 & 0 & 0 \end{array}
\right)^{\mathstrut}_{\mathstrut}$ 
& $\begin{array}{@{\,}c@{\,}}
\epsilon_1=\frac{-1}{8\pi} \frac{a_1^2 a_2^2}{a_1^2+b_1^2+d_1^2} 
\sin (2\xi) f(M_2^2/M_1^2)
\\
\epsilon_2=\frac{1^{\mathstrut}}{8\pi} a_1^2 \sin (2\xi) f(M_1^2/M_2^2) 
\end{array}$
\\ \hline
\end{tabular}
\label{table6}
\end{table}

\begin{table}
\caption{The allowed values of the parameters in $M_D$ and 
the predicted values of $\sin^2 \theta_{23}$, $\sin^2 \theta_{12}$, 
$\Delta m^2_{32}$, $\Delta m^2_{21}$ and 
$\langle m_{ee} \rangle$ in the cases $E$e5 and $E$e6.} 
\vspace{0.3cm}
\setlength{\tabcolsep}{4pt}\footnotesize
\begin{center}
\begin{tabular}{|c|c|c|} \hline
& $E$e5 & $E$e6   
\\ \hline
$a_1$ & $0.46-0.59$ & $0.019-0.023$
\\ \hline
$a_2$ & $0.17-0.24$ & $6.6-8.8$
\\ \hline
$b_2$ & $1.2-1.8$ & $1.4-2.0$
\\ \hline
$d_2$ & $1.1-1.5$ & $1.1-1.9$
\\ \hline
$\xi$ & $\sim \pi/2,3\pi/2$ & $\sim \pi/5,5\pi/6,5\pi/4,20\pi/11$ 
\\ \hline \hline
$\sin^2 \theta_{23}$ & $0.34-0.49$ & $0.34-0.49$ 
\\ \hline
$\sin^2 \theta_{12}$ & $0.26-0.40$ & $0.26-0.39$ 
\\ \hline
$\Delta m^2_{32}$ & $2.2 \times 10^{-3}-2.8 \times 10^{-3}$ 
& $2.0 \times 10^{-3}-2.8 \times 10^{-3}$ 
\\ \hline
$\Delta m^2_{21}$ & $7.2 \times 10^{-5}-8.3 \times 10^{-5}$ 
& $7.1 \times 10^{-5}-8.3 \times 10^{-5}$ 
\\ \hline
$\langle m_{ee} \rangle$ & $0.046-0.052$ 
& $0.044-0.052$
\\ \hline
\end{tabular}
\end{center}
\label{table7}
\end{table}

\section{Summary}

In this study, we have classified the Dirac and the right-handed Majorana 
neutrino mass matrices 
which can satisfy the SSA within the framework of the seesaw mechanism,
assuming that the Dirac neutrino mass matrix has some texture zeros. 
We found that the resulting Dirac neutrino mass matrices have rank 2 
as well as the rank of the effective neutrino mass matrix, or rank 1, 
depending only on the textures of $M_R^{-1}$.
We also considered the three cases of breaking the SSA by introducing a complex 
breaking parameter in the neutrino mass matrix and 
examined the effects of the breaking of the SSA on $|U_{e3}|, m_3$ and $J_{CP}$.

We have calculated the baryon asymmetry of the universe 
in the cases $E$e5 and $E$e6 which satisfy the SSA in the basis 
where $M_R$ is diagonal.
The implications of the baryon asymmetry for both cases are almost same.
We have also discussed the implications of the baryon asymmetry 
in the case of the breaking of the SSA for the cases $E$e5 and $E$e6.
We have found that only in the case $E$e6 with $b_3 \neq 0$ ($d_3 \neq 0$)
corresponding to the case A2 (A1), the CP asymmetry parameter 
$\epsilon_1$ can receive the correction from the breaking of the SSA.  
From the constraint of the observed value of $\eta_B$,
the order of magnitude of the breaking parameter $\epsilon$ should be of
the order of
$|\epsilon| \simeq \mathcal{O}(10^{-5})
\sim \mathcal{O}(10^{-3})$,
which can only generate the very small values of 
$m_3 \sim \mathcal{O}(10^{-9})-\mathcal{O}(10^{-6})$, 
$|U_{e3}| \sim \mathcal{O}(10^{-8})-\mathcal{O}(10^{-6})$ and 
$J_{CP} \sim \mathcal{O}(10^{-9})-\mathcal{O}(10^{-7})$.
Thus, observation of $|U_{e3}|$ in the next generation of reactor and long-baseline neutrino experiments 
will exclude the case $E$e6 with $b_3 \neq 0$ ($d_3 \neq 0$) as well as the original case of the SSA.

\section*{Acknowledgements}

I would like to thank Prof. Z.z. Xing for discussion and encouragement 
in the earlier stage of this work.
I would also like to thank W. Chao and H. Zhang for discussions about semi-analytic diagonalization. 
I would also like to thank L. Shu and K. Matsuda for comments.

\appendix

\section{Corrections for mass eigenvalues and eigenstates}
\clean

In this appendix, we will list the explicite forms of the mass eigenvalues and 
mixing angles up to the next-leading and the leading order approximation 
of the diagonalization of Eq. (\ref{neutrino-mass0}), respectively. 
The leading and the next-leading order of corrections for mass eigenvalues 
$(m_n^2)^{(1)}$'s and $(m_n^2)^{(2)}$'s are given by
\bea
(m_n^2)^{(1)}
&=&\langle n^{(0)}| \delta \mathcal{M} |n^{(0)} \rangle \quad (n=1,2,3) \, , 
\label{leadingmass}
\\
(m_n^2)^{(2)}
&=&\sum_{k \neq n} \langle n^{(0)}| \delta \mathcal{M} |k^{(0)} \rangle 
\langle k^{(0)}| \delta \mathcal{M} |n^{(0)} \rangle \quad (n,k=1,2,3) \, ,
\label{next-leadingmass}
\eea
and the leading order of corrections for mass eigenstates $|n^{(1)} \rangle$'s 
are given by 
\bea
|n^{(1)} \rangle= \sum_{k \neq n} 
\frac{|k^{(0)} \rangle \langle k^{(0)}| \delta \mathcal{M} |n^{(0)} \rangle }
{(m_n^2)^{(0)}-(m_k^2)^{(0)} }  \quad (n,k=1,2,3) \, .
\label{eigenstates}
\eea
In the next three subsections, we will write down the corresponding expressions 
for Eqs.~(\ref{leadingmass}), (\ref{next-leadingmass}) and 
(\ref{eigenstates}) in the three cases A1, A2 and A3.

\subsection{The case A1}

In the case A1,
the correction of the mass eigenvalues for the leading and 
the next-leading order of the approximation 
$(m_n^2)^{(1)}$'s and $(m_3^2)^{(2)}$ are given as follows~\footnote{
As we can see in Eq.~(\ref{A1m3}),
the leading order of correction $(m_3^2)^{(1)}$ 
includes only the term of the order of $|\epsilon|^2$.
Thus, we need to take into account the corrections 
up to the next-leading order for $m_3^2$, which also includes 
the terms of the order of $|\epsilon|^2$.}:
\bea
(m_1^2)^{(1)} &=& 2 D^2 
\Biggr( 1+ \frac{2}{c^2} \Biggl) \cos (\phi-\varphi) |\epsilon|
\sin^2 \theta 
\nonumber \\
&-& 2\frac{BD}{\sqrt{1+\frac{1}{c^2}}} \Biggr( 1+ \frac{2}{c^2} \Biggl)
\cos \varphi |\epsilon|
\sin \theta \cos \theta 
\nonumber \\
&+& D^2 \Biggr( 1+\frac{1}{c^2}+\frac{1}{1+c^2} \Biggl) 
|\epsilon|^2 \sin^2 \theta \, ,
\\
(m_2^2)^{(1)} &=& 2 D^2 
\Biggr( 1+ \frac{2}{c^2} \Biggl) \cos (\phi-\varphi) |\epsilon|
\cos^2 \theta 
\nonumber \\
&+& 2\frac{BD}{\sqrt{1+\frac{1}{c^2}}} \Biggr( 1+ \frac{2}{c^2} \Biggl)
\cos \varphi |\epsilon|
\sin \theta \cos \theta 
\nonumber \\
&+& D^2 \Biggr( 1+\frac{1}{c^2}+\frac{1}{1+c^2} \Biggl) 
|\epsilon|^2 \cos^2 \theta \, ,
\\
(m_3^2)^{(1)} &=& \frac{D^2}{c^2(1+c^2)} |\epsilon|^2 \, ,
\label{A1m3}
\\
(m_3^2)^{(2)} &=& \frac{-1}{A'D'(1+c^2)} 
\Biggl[ \frac{B^2D^2D'}{c^4} |\epsilon|^2
\nonumber \\
&+& A'D^4 |\epsilon|^2 
\Biggr( \frac{1}{c^4} +\frac{2\cos (\phi-\varphi)}{c^2(1+c^2)} |\epsilon|
+\frac{1}{(1+c^2)^2}|\epsilon|^2 \Biggl)
\nonumber \\
&+& 2\frac{BD^3B'}{c^2} |\epsilon|^2 
\Biggr( \frac{\cos \varphi}{c^2}
+\frac{\cos (\phi'-\phi+\varphi)}{(1+c^2)} |\epsilon| \Biggl)
\Biggr] \, .
\eea
The forms of $\langle k^{(0)}| \delta \mathcal{M} |n^{(0)} \rangle$'s 
in Eq.~(\ref{eigenstates}) can be expressed as follows:
\bea
\langle 3^{(0)}| \delta \mathcal{M} |1^{(0)} \rangle &=&
\frac{D^2}{c} 
\Biggr( \frac{e^{-i (\phi-\varphi)}}{c^2} +\frac{1}{1+c^2}|\epsilon| \Biggl) 
|\epsilon| \sin \theta
\nonumber \\
&-& \frac{BD}{c^2 \sqrt{1+c^2}} 
e^{i(\phi'-\phi+\varphi)} |\epsilon| \cos \theta
\\
&=& \langle 1^{(0)}| \delta \mathcal{M} |3^{(0)} \rangle^* \, ,
\\
\langle 3^{(0)}| \delta \mathcal{M} |2^{(0)} \rangle &=&
-\frac{D^2}{c} 
\Biggr( \frac{e^{-i (\phi-\varphi)}}{c^2} +\frac{1}{1+c^2}|\epsilon| \Biggl) 
|\epsilon| \cos \theta
\nonumber \\
&-& \frac{BD}{c^2 \sqrt{1+c^2}} 
e^{i(\phi'-\phi+\varphi)} |\epsilon| \sin \theta
\\
&=& \langle 2^{(0)}| \delta \mathcal{M} |3^{(0)} \rangle^* \, ,
\\
\langle 2^{(0)}| \delta \mathcal{M} |1^{(0)} \rangle &=&
-2 \frac{D^2}{c^2} 
\Biggr( 1+ \frac{2}{c^2} \Biggl) \cos (\phi-\varphi) |\epsilon|
\sin \theta \cos \theta
\nonumber \\
&+& \frac{BD}{\sqrt{1+\frac{1}{c^2}}} \Biggr( 1+ \frac{2}{c^2} \Biggl)
|\epsilon| (2\cos (\phi'-\phi+\varphi) \cos^2 \theta 
-e^{-i (\phi'-\phi+\varphi)})
\nonumber \\
&-& D^2 \Biggr( 1+ \frac{1}{c^2} \Biggl) |\epsilon|^2 \sin \theta \cos \theta
\\
&=& \langle 1^{(0)}| \delta \mathcal{M} |2^{(0)} \rangle^* \, .
\eea
Using Eqs. (\ref{sin}) and (\ref{cos}), 
we can obtain the correction of the three mixing angles for neutrinos as
\bea
(U_{\nu}^{(1)})_{e3} &=& -\frac{1}{A'D' \sqrt{1+\frac{1}{c^2}}} 
\nonumber \\
&\times& \Biggl[ 
\frac{D^2|B'|}{c} e^{i\phi'}
\Biggr( \frac{e^{-i (\phi-\varphi)}}{c^2} +\frac{1}{1+c^2}|\epsilon| \Biggl) 
|\epsilon| 
- \frac{BDD'}{c^3} 
e^{i(\phi-\varphi)} |\epsilon|  
\Biggr] \, , \\
(U_{\nu}^{(1)})_{\mu3} &=& \frac{1}{A'D'(1+\frac{1}{c^2})^{3/2}}
\nonumber \\
&\times& \Biggl[ 
\frac{A'D^2}{c}  
\Biggl( \frac{e^{-i (\phi-\varphi)}}{c^2}+\frac{1}{1+c^2} 
|\epsilon| \Biggl) |\epsilon| 
-\frac{BD|B'|}{c^3} e^{-i(\phi'-\phi+\varphi)} |\epsilon| 
\Biggr] \, , \\
(U_{\nu}^{(1)})_{e2} &=& \frac{e^{i \phi'} \cos \theta}{w} 
\Biggl[
-2 \frac{D^2}{c^2} 
\Biggr( 1+ \frac{2}{c^2} \Biggl) \cos (\phi-\varphi) |\epsilon|
\sin \theta \cos \theta
\nonumber \\
&+& \frac{BD}{\sqrt{1+\frac{1}{c^2}}} \Biggr( 1+ \frac{2}{c^2} \Biggl)
|\epsilon| (2\cos (\phi'-\phi+\varphi) \cos^2 \theta 
-e^{i (\phi'-\phi+\varphi)})
\nonumber \\
&-& D^2 \Biggr( 1+ \frac{1}{c^2} \Biggl) |\epsilon|^2 \sin \theta \cos \theta
\Biggr] \, .
\eea

\subsection{The case A2}

In the case A2, 
the correction of the mass eigenvalues for the leading and 
the next-leading order of the approximation 
$(m_n^2)^{(1)}$'s and $(m_3^2)^{(2)}$ are given as follows~\footnote{
Similarly to the case A1,
the leading order of correction $(m_3^2)^{(1)}$ 
includes only the term of the order of $|\epsilon|^2$ in Eq.~(\ref{A2m3}).
Thus, we need to take into account the corrections 
up to the next-leading order for $m_3^2$, which also includes 
the terms of the order of $|\epsilon|^2$.}:
\bea
(m_1^2)^{(1)} &=& D^2 
\Biggr( 2 \cos \varphi + \frac{c^2}{1+c^2}|\epsilon| \Biggl) |\epsilon|
\sin^2 \theta 
\nonumber \\
&-& 2\frac{BD}{\sqrt{1+\frac{1}{c^2}}} \cos (\phi'-\phi+\varphi) |\epsilon|
\sin \theta \cos \theta \, ,
\\
(m_2^2)^{(1)} &=& D^2 
\Biggr( 2 \cos \varphi + \frac{c^2}{1+c^2}|\epsilon| \Biggl) |\epsilon|
\cos^2 \theta 
\nonumber \\
&+& 2\frac{BD}{\sqrt{1+\frac{1}{c^2}}} \cos (\phi'-\phi+\varphi) |\epsilon|
\sin \theta \cos \theta \, ,
\\
(m_3^2)^{(1)} &=& \frac{D^2}{1+c^2} |\epsilon|^2 \, ,
\label{A2m3}
\\
(m_3^2)^{(2)} &=& \frac{-1}{A'D'(1+\frac{1}{c^2})} 
\Biggl[ \frac{B^2D^2D'}{c^2} |\epsilon|^2
+ A'D^4 |\epsilon|^2 
\Biggr( \frac{1}{c^2} +\frac{2\cos \varphi}{1+c^2} |\epsilon|
+\frac{c^2}{(1+c^2)^2}|\epsilon|^2 \Biggl)
\nonumber \\
&-& 2\frac{BD^3B'}{c^2} |\epsilon|^2 
\Biggr( \cos (\phi'+\varphi)
+\frac{c^2}{(1+c^2)^2} \cos (\phi'-\phi+\varphi)|\epsilon| \Biggl)
\Biggr] \, .
\eea
The forms of $\langle k^{(0)}| \delta \mathcal{M} |n^{(0)} \rangle$'s 
in Eq. (\ref{eigenstates}) can be expressed as follows:
\bea
\langle 3^{(0)}| \delta \mathcal{M} |1^{(0)} \rangle &=&
\frac{D^2}{c} \Biggr( e^{i \varphi} +\frac{c^2}{1+c^2}|\epsilon| \Biggl) 
|\epsilon| \sin \theta
\nonumber \\
&-& \frac{BD}{\sqrt{1+c^2}} e^{i(\phi'-\phi+\varphi)} |\epsilon| \cos \theta
\\
&=& \langle 1^{(0)}| \delta \mathcal{M} |3^{(0)} \rangle^* \, ,
\\
\langle 3^{(0)}| \delta \mathcal{M} |2^{(0)} \rangle &=&
-\frac{D^2}{c} \Biggr( e^{i \varphi} +\frac{c^2}{1+c^2}|\epsilon| \Biggl) 
|\epsilon| \cos \theta
\nonumber \\
&-& \frac{BD}{\sqrt{1+c^2}} e^{i(\phi'-\phi+\varphi)} |\epsilon| \sin \theta
\\
&=& \langle 2^{(0)}| \delta \mathcal{M} |3^{(0)} \rangle^* \, ,
\\
\langle 2^{(0)}| \delta \mathcal{M} |1^{(0)} \rangle &=&
-D^2 \Biggr( 2 \cos \varphi +\frac{c^2}{1+c^2}|\epsilon| \Biggl) 
|\epsilon| \sin \theta \cos \theta
\nonumber \\
&+& \frac{BD}{\sqrt{1+\frac{1}{c^2}}} |\epsilon| 
(2\cos (\phi'-\phi+\varphi) \cos^2 \theta-e^{-i(\phi'-\phi+\varphi)})
\\
&=& \langle 1^{(0)}| \delta \mathcal{M} |2^{(0)} \rangle^* \, .
\eea
Using Eqs. (\ref{sin}) and (\ref{cos}), 
we can obtain the correction of the three mixing angles for neutrinos as
\bea
(U_{\nu}^{(1)})_{e3} &=& \frac{-1}{A'D' \sqrt{1+\frac{1}{c^2}}} 
\nonumber \\
&\times& \Biggl[ 
\frac{D^2 |B'|}{c} e^{i\phi'} 
\Biggl( e^{-i\varphi}+\frac{c^2}{1+c^2} |\epsilon| \Biggl) |\epsilon|
-\frac{BDD'}{c} e^{i(\phi-\varphi)} |\epsilon| 
\Biggr] \, ,
\\
(U_{\nu}^{(1)})_{\mu3} &=& \frac{1}{A'D'(1+\frac{1}{c^2})^{3/2}}
\nonumber \\
&\times& \Biggl[ 
\frac{A'D^2}{c}  
\Biggl( e^{-i\varphi}+\frac{c^2}{1+c^2} |\epsilon| \Biggl) |\epsilon| 
-\frac{BD|B'|}{c} e^{-i(\phi'-\phi+\varphi)} |\epsilon| 
\Biggr] \, ,
\\
(U_{\nu}^{(1)})_{e2} &=& \frac{e^{i \phi'} \cos \theta}{w} 
\Biggl[
-D^2 \Biggr( 2 \cos \varphi +\frac{c^2}{1+c^2}|\epsilon| \Biggl) 
|\epsilon| \sin \theta \cos \theta
\nonumber \\
&+& \frac{BD}{\sqrt{1+\frac{1}{c^2}}} |\epsilon| 
(2\cos (\phi'-\phi+\varphi) \cos^2 \theta-e^{i(\phi'-\phi+\varphi)})
\Biggr] \, .
\eea


\subsection{The case A3}

In the case A3, 
the correction of the mass eigenvalues for the leading and 
the next-leading order of the approximation 
$(m_n^2)^{(1)}$'s and $(m_3^2)^{(2)}$ are given as follows~\footnote{
Similarly to the cases A1 and A2,
the leading order of correction $(m_3^2)^{(1)}$ given in Eq.~(\ref{A3m3}) 
includes only the term of the order of $|\epsilon|^2$.
Thus, we need to take into account the next-leading order of corrections, 
which also includes the terms of the order of $|\epsilon|^2$.
However, the terms of the order of $|\epsilon|^2$ in the leading and 
the next-leading order corrections almost cancel.
Therefore, we need to take into account the higer order corrections for $m_3^2$ 
in the case A3. 
Here we do not write down the expressions. 
We checked the results for $m_3^2$ up to the forth order, 
comparing with the numerical calculation of the diagonalization and 
found that this perturbation is not good for $m_3^2$ in the case A3.}:
\bea
(m_1^2)^{(1)} &=& 2B^2 |\epsilon| \cos^2 \theta
+2B^2 |\epsilon| \cos \varphi \sin^2 \theta
\nonumber \\
&-& 2BD \sqrt{1+\frac{1}{c^2}} \cos (\phi'-\phi'') |\epsilon| 
\sin \theta \cos \theta 
\nonumber \\
&-& 2\frac{AB}{\sqrt{1+\frac{1}{c^2}}} \cos (\phi'+\phi'') |\epsilon|
\sin \theta \cos \theta 
\nonumber \\
&+& B^2 \frac{c^2}{1+c^2} |\epsilon|^2 \sin^2 \theta 
+ B^2 |\epsilon|^2 \cos^2 \theta \, ,
\\
(m_2^2)^{(1)} &=& 2B^2 |\epsilon| \sin^2 \theta
+2B^2 |\epsilon| \cos \varphi \cos^2 \theta 
\nonumber \\
&+& 2BD \sqrt{1+\frac{1}{c^2}} \cos (\phi'-\phi'') |\epsilon| 
\sin \theta \cos \theta 
\nonumber \\
&+& 2\frac{AB}{\sqrt{1+\frac{1}{c^2}}} \cos (\phi'+\phi'') |\epsilon|
\sin \theta \cos \theta
\nonumber \\
&+& B^2 \frac{c^2}{1+c^2} |\epsilon|^2 \cos^2 \theta 
+B^2 |\epsilon|^2 \sin^2 \theta \, ,
\\
(m_3^2)^{(1)} &=& \frac{B^2}{1+c^2} |\epsilon|^2 \, ,
\label{A3m3}
\\
(m_3^2)^{(2)} &=& \frac{-1}{A'D'(1+\frac{1}{c^2})} 
\Biggl[ \frac{A^2 B^2 D'}{c^2} |\epsilon|^2+A'B^4 |\epsilon|^2
\Biggr( \frac{1}{c^2} +\frac{2\cos \varphi}{1+c^2} |\epsilon|
+\frac{c^2}{(1+c^2)^2}|\epsilon|^2 \Biggl) 
\nonumber \\
&-& 2\frac{AB^3B'}{c^2} |\epsilon|^2 
\Biggr( \cos (\phi'+\varphi)
+\frac{c^2}{1+c^2} \cos (\phi'-\phi+\varphi)|\epsilon| \Biggl)
\Biggr] \, ,
\eea 
where $\phi''=\phi+\varphi$.

The forms of $\langle k^{(0)}| \delta \mathcal{M} |n^{(0)} \rangle$'s 
in Eq. (\ref{eigenstates}) can be expressed as follows:
\bea
\langle 3^{(0)}| \delta \mathcal{M} |1^{(0)} \rangle &=&
\frac{B^2}{c} \Biggr( e^{i \varphi}+\frac{c^2}{1+c^2}|\epsilon| \Biggl) 
|\epsilon| \sin \theta
\nonumber \\
&-& \frac{AB}{\sqrt{1+c^2}} e^{i(\phi'+\phi'')} |\epsilon| \cos \theta
\\
&=& \langle 1^{(0)}| \delta \mathcal{M} |3^{(0)} \rangle^* \, ,
\\
\langle 3^{(0)}| \delta \mathcal{M} |2^{(0)} \rangle &=&
-\frac{B^2}{c} \Biggr( e^{i \varphi}+\frac{c^2}{1+c^2}|\epsilon| \Biggl) 
|\epsilon| \cos \theta
\nonumber \\
&-& \frac{AB}{\sqrt{1+c^2}} e^{i(\phi'+\phi'')} |\epsilon| \sin \theta
\\
&=& \langle 2^{(0)}| \delta \mathcal{M} |3^{(0)} \rangle^* \, ,
\\
\langle 2^{(0)}| \delta \mathcal{M} |1^{(0)} \rangle &=&
2B^2 (1-\cos \varphi) |\epsilon| \sin \theta \cos \theta
+B^2 \frac{1}{1+c^2} |\epsilon|^2 \sin \theta \cos \theta
\nonumber \\
&+& BD \sqrt{1+\frac{1}{c^2}} |\epsilon| 
(2\cos (\phi'-\phi'') \cos^2 \theta-e^{-i(\phi'-\phi'')})
\nonumber \\
&+&\frac{AB}{\sqrt{1+\frac{1}{c^2}}} |\epsilon| 
(2\cos (\phi'+\phi'') \cos^2 \theta-e^{-i(\phi'+\phi'')})
\\
&=& \langle 1^{(0)}| \delta \mathcal{M} |2^{(0)} \rangle^* \, .
\eea
Using Eqs. (\ref{sin}) and (\ref{cos}), 
we can obtain the correction of the three mixing angles for neutrinos as
\bea
(U_{\nu}^{(1)})_{e3} &=& \frac{1}{A'D' \sqrt{1+\frac{1}{c^2}}} 
\nonumber \\
&\times& \Biggl[ 
-\frac{B^2 |B'|}{c} e^{i\phi'} 
\Biggl( e^{-i\varphi} +\frac{c^2}{1+c^2} |\epsilon| \Biggl) |\epsilon|
+\frac{ABD'}{c} e^{-i\phi''} |\epsilon| 
\Biggr] \, ,
\\
(U_{\nu}^{(1)})_{\mu3} &=& \frac{1}{A'D'(1+\frac{1}{c^2})^{3/2}}
\nonumber \\
&\times& \Biggl[ 
\frac{A'B^2}{c}  
\Biggl( e^{-i\varphi} +\frac{c^2}{1+c^2} |\epsilon| \Biggl) |\epsilon|
-\frac{AB|B'|}{c} e^{-i(\phi'+\phi'')} |\epsilon| 
\Biggr] \, ,
\\
(U_{\nu}^{(1)})_{e2} &=& \frac{e^{i \phi'} \cos \theta}{w}
\Biggl[ 
2B^2 (1-\cos \varphi) |\epsilon| \sin \theta \cos \theta
+B^2 \frac{1}{1+c^2} |\epsilon|^2 \sin \theta \cos \theta
\nonumber \\
&+& BD \sqrt{1+\frac{1}{c^2}} |\epsilon| 
(2\cos (\phi'-\phi'') \cos^2 \theta-e^{i(\phi'-\phi'')})
\nonumber \\
&+&\frac{AB}{\sqrt{1+\frac{1}{c^2}}} |\epsilon| 
(2\cos (\phi'+\phi'') \cos^2 \theta-e^{i(\phi'+\phi'')})
\Biggr] \, .
\eea


\end{document}